\documentclass[
    pra,                
    aps,                 
    twocolumn,             
    groupedaddress,     
    longbibliography,       
    nofootinbib,           
    floatfix,
    10pt               
]{revtex4-2}

\usepackage[utf8]{inputenc}
\usepackage[T1]{fontenc}  
\usepackage{microtype}     
\microtypecontext{spacing=nonfrench}

\usepackage{amsmath}
\usepackage{amssymb} 
\usepackage{amsfonts}
\usepackage{mathrsfs}
\usepackage{bm}

\usepackage{amsthm} 

\usepackage{graphicx} 

\usepackage{booktabs} 
\usepackage{multirow}
\usepackage{array}    
\usepackage{xcolor}  
\usepackage{colortbl} 
\usepackage{makecell} 

\usepackage{textcomp}      

\usepackage{siunitx}
\usepackage[colorlinks=true,
            linkcolor=blue,       
            urlcolor=blue,        
            citecolor=blue,       
            anchorcolor=blue,
            breaklinks=true]{hyperref} 
\usepackage[capitalize]{cleveref}
\usepackage[dvipsnames]{xcolor}

\usepackage[ruled,vlined,linesnumbered]{algorithm2e}

\DeclareMathOperator*{\argmin}{arg\,min}

\begin{document}

\newcommand{\nudt}{{College of Computer Science and Technology, National University of Defense Technology, Changsha 410073, China}}
\author{Yuanqi Liu} \affiliation{\nudt}
\author{Weilei Zeng}\affiliation{\nudt}
\author{Peixiang Li} \affiliation{\nudt}
\author{Yantong Liu} \affiliation{\nudt}
\author{Guangyao Huang} \affiliation{\nudt}
\author{Yingwen Liu} \affiliation{\nudt}
\author{Dongyang Wang} \affiliation{\nudt}
\author{Junjie Wu} \affiliation{\nudt}
\author{Lingling Lao} \email{laolinglingrolls@gmail.com}\affiliation{\nudt}
\date{\today}

\title{An iterative Ising decoder for quantum error correction codes}

\begin{abstract}
The Ising framework maps the decoding problem in quantum error correction onto ground-state optimization of a classical Hamiltonian, in which $X$-$Z$ error correlations enter as cross terms. 
Under phenomenological depolarizing noise, the exact joint formulation contains up to 8-body interactions for the toric code and 10-body for the $6.6.6$ color code. These high-order terms degrade solver convergence, inflate runtime, and raise the auxiliary spin overhead when embedding into native 2-body Ising hardware.
In this work, we propose the iterative low-order decoding (ILOD) algorithm, which alternates between $X$- and $Z$-type sub-Hamiltonians, approximating cross-type correlations through Bayesian priors that reweight each type's couplings using the other type's inferred error configuration.
This halves the maximum body count of interaction terms in the Hamiltonian, accelerating the solver, restoring convergence at larger code distances, and reducing the total spin count for 2-body embedding by a factor of $2.5$. 
For the toric code, ILOD attains a threshold of $4.73\%$ versus $4.83\%$ for the joint formulation, with the empirical runtime ratio scaling as $(0.81)^d$. For the $6.6.6$ color code, their thresholds agree within statistical uncertainty for small code distances, and ILOD remains convergent for larger distances where the joint formulation fails to converge despite a larger annealing budget.
\end{abstract}
 
\maketitle

\section{INTRODUCTION}
\label{sec:introduction}

In a fault-tolerant quantum computation architecture, the decoder determines recovery operations from syndrome measurements within the classical control stack~\cite{terhal2015,battistel2023real}.
A practical decoder must balance three competing objectives: high accuracy (suppressing the logical error rate toward the optimal decoding limit), low latency (avoiding data backlog), and scalability (maintaining tractable computational and hardware overhead as the code distance grows)~\cite{Fowler2012,terhal2015}. 
The Ising framework approaches decoding through statistical mechanics~\cite{sourlas_spin-glass_1989,Dennis2002,Katzgraber2009,Bombin2012} and can achieve near-optimal logical error rates by naturally modeling complex error correlations and code topology.
Specifically, it maps the decoding problem onto ground-state optimization of a classical random-bond Ising-type Hamiltonian, where syndrome information is encoded as quenched disorder in the couplings. The ground state identifies the single most probable error configuration, implementing minimum-weight decoding (MWD).
For bit-flip noise on the toric code, this mapping yields a 2-body Hamiltonian directly solvable on dedicated Ising hardware, such as coherent Ising machines~\cite{Marandi2014,Inagaki2016}, quantum annealers~\cite{Johnson2011,King2023}, and digital annealers~\cite{Aramon2019}.
These platforms have scaled to 2-body instances of up to $10^5$ spins with sub-millisecond time-to-solution~\cite{honjo_100000_spin_2021}, motivating native 2-body Ising hardware as a candidate substrate for latency-critical decoding.

Beyond bit-flip errors, the Ising framework captures the $X$-$Z$ error correlations induced by $Y$ errors through cross terms in the Hamiltonian.
Under code-capacity depolarizing noise, however, these cross terms increase the maximum body count of interaction terms from 2- to 4-body for the toric code~\cite{takeuchi2023depolarizingising} and from 3- to 6-body for the $6.6.6$ color code~\cite{Katzgraber2009,Bombin2012}. 
Extending the spacetime stabilizer parametrization~\cite{takada2024colorising} from phenomenological bit-flip noise to the depolarizing case further raises these body counts to 8 and 10, respectively.
Such high-order interactions could create rugged energy landscapes that hinder heuristic ground-state searches, degrading solver convergence and inflating runtime.
While the proliferation of metastable states is analytically established for fully connected spin-glass models~\cite{derrida1981random,gross1984simplest}, analogous bottlenecks are expected on the sparse topological graphs considered here.
These solver bottlenecks are compounded by the embedding cost: because mainstream Ising hardware natively supports only 2-body interactions, deploying these high-order terms requires many auxiliary spins per 8- and 10-body term~\cite{Rosenberg1975,Boros2002,dattani2019quadratization}, with the cumulative overhead growing rapidly in the interaction order and posing a scalability challenge as the code distance grows.

To address these bottlenecks, we propose the iterative low-order decoding (ILOD) algorithm. Rather than optimizing the joint high-order Hamiltonian, ILOD separates the optimization into $X$- and $Z$-type sub-Hamiltonians solved in alternation, which halves the maximum body count of interaction terms from 8- and 10-body to 4- and 5-body for the toric and color codes, respectively. Cross-type correlations are approximately captured through Bayesian priors that reweight the $X$-type ($Z$-type) couplings using the error configuration inferred from the $Z$ type ($X$ type). This body-count reduction yields decoding gains in solver acceleration, restored convergence at larger code distances, and lower spin overhead after quadratization (the conversion of high-order interactions to 2-body terms via auxiliary spins).
Numerical simulations under phenomenological depolarizing noise show that, for the toric code, ILOD attains a threshold of $4.73\%$ versus $4.83\%$ for the joint formulation, both exceeding minimum-weight perfect matching (MWPM, $3.80\%$) and its correlation-aware variant (Corr-MWPM, $4.65\%$), with the empirical runtime ratio (ILOD/joint) scaling as $(0.81)^d$ over the code distance $d \in \{3,5,7,9\}$. For the $6.6.6$ color code, ILOD achieves a threshold of $4.41\%$, agreeing with the joint formulation's $4.36\%$ within statistical uncertainty, and additionally restores solver convergence at $d=9$, where the joint formulation fails to converge despite a larger annealing budget. ILOD also enables a $2.5\times$ reduction in the total spin count for both codes after conversion to native 2-body forms, lowering the resource requirement on dedicated Ising hardware.

The remainder of this paper is organized as follows. Section~\ref{sec:preliminaries} reviews the stabilizer formalism and develops the Ising mapping under bit-flip, depolarizing, and phenomenological noise. Section~\ref{sec:ilod} presents the ILOD algorithm. Section~\ref{sec:results} reports numerical results. Section~\ref{sec:conclusion_discussion} discusses limitations and hardware deployment prospects, and concludes the paper.

\section{PRELIMINARIES}
\label{sec:preliminaries}

\subsection{Quantum error correction and stabilizer codes}
\label{sec:stabilizer_codes}

Stabilizer codes are defined by abelian subgroups $\mathcal{S} \subset \mathcal{G}_n$ such that $-I \notin \mathcal{S}$, where  $\mathcal{G}_n$ represents the $n$-qubit Pauli group, consisting of $n$-fold tensor products of $\{I, X, Y, Z\}$ with phases $\{\pm 1, \pm \mathrm{i}\}$. An $[[n, k, d]]$ stabilizer code encodes $k$ logical qubits into $n$ physical qubits. Given $r = n-k$ independent generators $\{g_1, \ldots, g_r\}$ of $\mathcal{S}$, the code space $\mathcal{C}$ is their simultaneous $+1$ eigenspace:
\begin{equation}
    \mathcal{C} = \bigl\{ |\psi\rangle \in (\mathbb{C}^2)^{\otimes n} : g_i |\psi\rangle = |\psi\rangle, \; \forall i = 1, \ldots, r \bigr\}.
    \label{eq:code_space}
\end{equation}
The code distance $d$ is the minimum Pauli weight of an operator in $N(\mathcal{S}) \setminus \mathcal{S}$, where $N(\mathcal{S})$ is the normalizer of $\mathcal{S}$ in $\mathcal{G}_n$. Operators in this set, termed logical operators, act non-trivially on $\mathcal{C}$ without stabilizing it; two such operators differing by an element of $\mathcal{S}$ act identically on $\mathcal{C}$, so the cosets $N(\mathcal{S})/\mathcal{S}$ partition $N(\mathcal{S})$ into \emph{logical classes} (equivalently, logical cosets).

For a Pauli error $E \in \mathcal{G}_n$, measuring each generator $g_i$ yields a binary syndrome vector $\boldsymbol{\gamma} \in \{0,1\}^r$, where
\begin{equation}
    \gamma_i = 
    \begin{cases} 
        0 & \text{if } [g_i, E] = 0, \\
        1 & \text{if } \{g_i, E\} = 0.
    \end{cases}
    \label{eq:syndrome}
\end{equation}
Since different error configurations can cause the same syndrome, the decoding task simplifies to identifying the most likely logical class rather than any specific representative~\cite{Poulin2006}.
To formalize this equivalence class structure, we identify each Pauli error configuration $E$ with its binary symplectic representation $\boldsymbol{e} = (\boldsymbol{e}^X, \boldsymbol{e}^Z) \in \{0,1\}^{2n}$, where $e^X_q, e^Z_q \in \{0,1\}$ encode the $X$ and $Z$ supports on qubit $q$. Let $S \in \{0,1\}^{r \times 2n}$ be the parity-check matrix of $\mathcal{S}$, $L \in \{0,1\}^{2k \times 2n}$ the logical generator matrix, and $D \in \{0,1\}^{r \times 2n}$ the destabilizer matrix defined by $S \Lambda D^{T} = I_{r} \pmod{2}$, where $\Lambda = \bigl(\begin{smallmatrix} 0 & I_n \\ I_n & 0 \end{smallmatrix}\bigr)$ is the symplectic form~\cite{gottesman1997stabilizer,Aaronson2004}. The rows of $D$, $S$, and $L$ together form a symplectic basis of $\mathbb{F}_2^{2n}$, so any error decomposes uniquely as
\begin{equation}
    \boldsymbol{e} = \boldsymbol{\gamma} D + \boldsymbol{\alpha} S + \boldsymbol{\ell} L \pmod 2.
    \label{eq:error-configuration}
\end{equation}
Each row of $D$ anticommutes with exactly one stabilizer generator, so $\boldsymbol{\epsilon} \equiv \boldsymbol{\gamma} D$ is a \emph{pure error} consistent with the measured syndrome $\boldsymbol{\gamma}$. The vector $\boldsymbol{\alpha} \in \{0,1\}^{r}$ parameterizes the stabilizer degrees of freedom within an equivalence class, while $\boldsymbol{\ell} \in \mathcal{L} \equiv \{0,1\}^{2k}$ labels the class itself. As $\boldsymbol{\alpha}$ and $\boldsymbol{\ell}$ vary, Eq.~\eqref{eq:error-configuration} enumerates all errors consistent with $\boldsymbol{\gamma}$; we write $\boldsymbol{e}(\boldsymbol{\gamma}, \boldsymbol{\alpha}, \boldsymbol{\ell})$ for this parametrization.

Under minimum-weight decoding (MWD), the decoder selects the logical class containing the globally minimum-weight error:
\begin{equation}
    \hat{\boldsymbol{\ell}}_{\mathrm{MWD}} = \argmin_{\boldsymbol{\ell} \in \mathcal{L}} \min_{\boldsymbol{\alpha} \in \{0,1\}^{r}} w\bigl(\boldsymbol{e}(\boldsymbol{\gamma}, \boldsymbol{\alpha}, \boldsymbol{\ell})\bigr),
    \label{eq:MWD}
\end{equation}
where $w(\cdot)$ denotes the symplectic Pauli weight, defined as the number of qubits on which the operator acts non-trivially: $w(\boldsymbol{e}) = \bigl|\bigl\{q \in \{1,\dots,n\} \mid e^X_q = 1 \text{ or } e^Z_q = 1\bigr\}\bigr|$.

Throughout this work, we focus on CSS codes~\cite{calderbankshor1996,steane1996}, where the parity-check matrix decomposes as $S = \begin{pmatrix} S^X & 0 \\ 0 & S^Z \end{pmatrix}$, with $r^X$ and $r^Z$ generators of $X$- and $Z$-type stabilizers, respectively ($r = r^X + r^Z$), yielding decoupled syndrome constraints. The class-by-class parametrization in Eq.~\eqref{eq:error-configuration}, specialized to the CSS structure, is the basis of the Ising mappings in Sec.~\ref{sec:mapping}.

\subsubsection{Toric codes}
\label{sec:toric_background}

The toric code~\cite{Dennis2002,Kitaev2003} is a topological stabilizer code defined on an $L \times L$ square lattice with periodic boundary conditions. It encodes $k = 2$ logical qubits into $n = 2L^2$ physical qubits with code distance $d = L$.

\begin{figure}[htbp]
    \centering
    \includegraphics[width=0.6\linewidth]{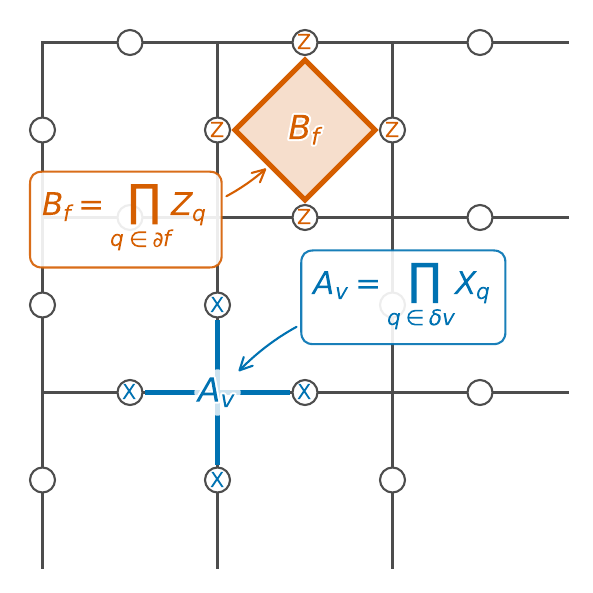}
    \caption{Geometry of the toric code. Data qubits (circles) reside on the lattice edges. Vertex operators $A_v$ (blue) and face operators $B_f$ (orange) represent the two types of weight-4 stabilizer generators.}
    \label{fig:toric_code}
\end{figure}

As illustrated in Fig.~\ref{fig:toric_code}, data qubits reside on the lattice edges. The stabilizer group is generated by vertex operators ($A_v$) and face operators ($B_f$), which act on the four edges incident to a vertex $v$ or bounding a face $f$, respectively:
\begin{equation}
    A_v = \prod_{q \in \delta v} X_q, \qquad
    B_f = \prod_{q \in \partial f} Z_q,
    \label{eq:toric_stabilizers}
\end{equation}
where $X_q$ and $Z_q$ denote Pauli operators on the $q$-th qubit, so that $A_v$ detects $Z$ errors and $B_f$ detects $X$ errors.
Because every data qubit resides on an edge incident to two vertices and shared by two faces, a single $X$ ($Z$) error violates exactly two $Z$-type ($X$-type) stabilizers. This connectivity yields a matching graph in which each data qubit corresponds to an edge connecting the two violated stabilizers.

\subsubsection{Color codes}
\label{sec:color_background}

The color code is a family of topological stabilizer codes defined on trivalent, three-face-colorable lattices. An example is the $6.6.6$ hexagonal lattice with a triangular boundary~\cite{Bombin2006}, encoding $k = 1$ logical qubit into $n = (3d^2+1)/4$ physical qubits for odd $d$, where $d$ equals the boundary side length.

\begin{figure}[tbp]
    \centering
    \includegraphics[width=0.6\linewidth]{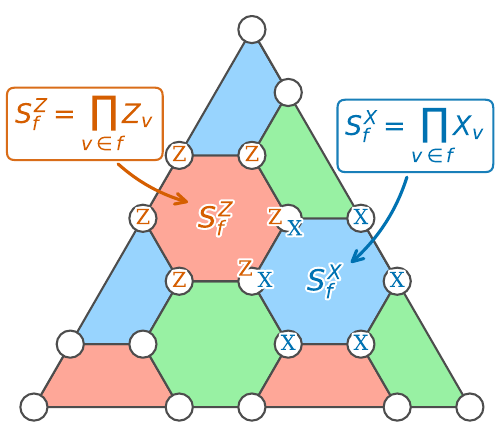}
    \caption{Geometry of the $6.6.6$ color code with a triangular boundary. Data qubits (circles) reside at the lattice vertices. Each bulk hexagonal face (red, green, or blue) defines weight-6 $X$- and $Z$-type stabilizer generators, while the truncated faces on the boundary define weight-4 generators.}
    \label{fig:color_code}
\end{figure}

As illustrated in Fig.~\ref{fig:color_code}, data qubits are located at the lattice vertices. Every face $f$ defines two stabilizer generators acting on its boundary vertices $\partial f$:
\begin{equation}
    S_f^X = \prod_{v \in \partial f} X_v, \qquad S_f^Z = \prod_{v \in \partial f} Z_v.
    \label{eq:color_stabilizers}
\end{equation}

Every bulk data qubit sits at a trivalent vertex incident to three faces, so a single $X$ ($Z$) error violates three $Z$-type ($X$-type) stabilizers (boundary qubits violate fewer). This connectivity induces a decoding hypergraph for the $X$- and $Z$-type syndromes separately, in which data qubits correspond to hyperedges of size at most $3$.

\subsection{Ising mapping for minimum-weight decoding}
\label{sec:mapping}

For the independent and identically distributed noise models considered in this work (bit-flip and depolarizing), the probability of an error configuration depends only on its symplectic Pauli weight $w(\boldsymbol{e})$. Consequently, the negative log-likelihood $-\ln P(\boldsymbol{e})$ is linear in the error weight $w(\boldsymbol{e})$, and minimizing $w$ in Eq.~\eqref{eq:MWD} is equivalent to minimizing $-\ln P(\boldsymbol{e})$.
With $H(\boldsymbol{e}) \equiv -\ln P(\boldsymbol{e})$ as a classical Hamiltonian, fixing the class $\boldsymbol{\ell}$ in Eq.~\eqref{eq:error-configuration} parametrizes the error solely by the stabilizer variable $\boldsymbol{\alpha}$, so that the inner minimization in Eq.~\eqref{eq:MWD} is reduced to a ground-state search of the class-conditional Hamiltonian $H_{\boldsymbol{\ell}}(\boldsymbol{\alpha}) \equiv H(\boldsymbol{e}(\boldsymbol{\gamma}, \boldsymbol{\alpha}, \boldsymbol{\ell}))$ in a disordered Ising model~\cite{sourlas_spin-glass_1989,Dennis2002,Nishimori1981,kovalev_spin_2015}. The MWD estimate $\hat{\boldsymbol{\ell}}_{\mathrm{MWD}}$ then follows by comparing the ground-state energies across all $|\mathcal{L}| = 2^{2k}$ logical classes. For CSS codes, this splits into independent $X$- and $Z$-type comparisons over $2^k$ classes each. 
The following subsections construct the class-conditional Hamiltonian $H_{\boldsymbol{\ell}}(\boldsymbol{\alpha})$ explicitly for three noise models of increasing complexity: bit-flip, depolarizing, and phenomenological.

\subsubsection{Bit-flip noise model}
\label{sec:bit_flip}

Under independent bit-flip noise, each qubit undergoes a Pauli-$X$ error with probability $p$. Since only $X$ errors occur, we retain only the $X$-component $\boldsymbol{e}^X \in \{0,1\}^n$ of the symplectic representation ($\boldsymbol{e}^Z \equiv 0$), and the distribution factorizes as
\begin{equation}
    P(\boldsymbol{e}^X) = \prod_{q=1}^{n} p^{e^X_q}(1-p)^{1-e^X_q},
    \label{eq:bitflip_prob}
\end{equation}
which takes the Boltzmann form $P(\boldsymbol{\tau}) \propto e^{-\beta H_0(\boldsymbol{\tau})}$ of a non-interacting paramagnet under the spin transformation $e^X_q = (1-\tau_q)/2$ with $\tau_q \in \{\pm 1\}$,
\begin{equation}
    H_0(\boldsymbol{\tau}) = -J \sum_{q=1}^n \tau_q, \qquad \beta J = \tfrac{1}{2}\ln\frac{1-p}{p},
    \label{eq:paramagnet}
\end{equation}
where $\beta$ is the inverse temperature and the Nishimori condition~\cite{Nishimori1981} fixes the product $\beta J$ in terms of $p$.

The Boltzmann form thus reduces decoding to a ground-state search of $H_0(\boldsymbol{\tau})$ over $X$-errors consistent with the measured syndrome. Specializing Eq.~\eqref{eq:error-configuration} to a fixed logical class (the class label will be omitted in the following) and restricting to the $X$-type, we write $\boldsymbol{e}^X = \boldsymbol{\epsilon}^X + \boldsymbol{\alpha}^X S^X \pmod 2$, where $\boldsymbol{\epsilon}^X$ is the $X$-component of the pure error $\boldsymbol{\gamma} D$, and $S^X \in \{0,1\}^{r^X \times n}$ is the parity-check matrix of the $X$-type stabilizers. Substituting this parametrization into $\tau_q = (-1)^{e^X_q}$ yields
\begin{equation}
    \tau_q = \eta_q \prod_{j=1}^{r^X} \sigma_j^{S^X_{j,q}},
    \label{eq:gauge_mapping}
\end{equation}
where $\eta_q \equiv (-1)^{\epsilon^X_q}$ is the quenched disorder fixed by the pure error, $\sigma_j \equiv (-1)^{\alpha^X_j} \in \{\pm 1\}$ is the gauge spin (representing the stabilizer degree of freedom) to be optimized, and $S^X_{j,q} \in \{0,1\}$ selects which $X$-stabilizers act non-trivially on qubit $q$.

Substituting Eq.~\eqref{eq:gauge_mapping} into Eq.~\eqref{eq:paramagnet} transforms the paramagnet into a random-bond Ising model (RBIM), where each qubit contributes a product of the gauge spins of the stabilizers acting on it:
\begin{equation}
    H(\boldsymbol{\sigma}) = -J \sum_{q=1}^{n} \eta_q \prod_{j=1}^{r^X} \sigma_j^{S^X_{j,q}}.
    \label{eq:rbim_general}
\end{equation}
Each gauge spin $\sigma_j$ corresponds to an $X$-stabilizer, each term in the sum corresponds to the $q$-th data qubit, and the product is over the set $N^X(q) \equiv \{j : S^X_{j,q} = 1\}$ of $X$-stabilizers incident to qubit $q$. The syndrome enters as quenched disorder through $\boldsymbol{\eta}$: a non-trivial pure-error component ($\eta_q = -1$) flips the sign of the $q$-th interaction term. The decoding problem is thereby reduced to finding the ground state of Eq.~\eqref{eq:rbim_general}.

For the toric code, each qubit is incident to two $X$-type vertices; placing a gauge spin at each vertex maps them to the primal square lattice. Labeling these vertices by $i, j$, the RBIM takes the two-body form
\begin{equation}
    H_{\mathrm{toric}}(\boldsymbol{\sigma}) = -J \sum_{\{i,j\} \in \mathcal{E}} \eta_{ij}\, \sigma_i \sigma_j,
    \label{eq:toric_hamiltonian}
\end{equation}
where $\mathcal{E}$ is the edge set of the lattice and each edge corresponds to one physical qubit. This reproduces the standard two-dimensional RBIM on the square lattice. For the $6.6.6$ color code, each bulk qubit lies in the support of three $X$-type stabilizers, and Eq.~\eqref{eq:rbim_general} translates to a three-body interaction:
\begin{equation}
    H_{\mathrm{color}}(\boldsymbol{\sigma}) = -J \sum_{e \in \mathcal{E}^*} \eta_{e} \prod_{v \in e} \sigma_v,
    \label{eq:color_hamiltonian}
\end{equation}
where $\mathcal{E}^*$ is the set of hyperedges corresponding to the physical data qubits. Geometrically, on the dual triangular lattice, bulk qubits define three-body interactions on triangular faces ($|e|=3$), while boundary qubits reduce to two-body interactions on boundary edges ($|e|=2$). Under bit-flip noise, the maximum body count of interaction terms is therefore $2$ for the toric code and $3$ for the $6.6.6$ color code.

\subsubsection{Depolarizing noise model}
\label{sec:depolarizing}

For depolarizing noise, each qubit independently undergoes an $X$, $Y$, or $Z$ error with probability $p/3$ each (and no error with probability $1-p$). Because depolarizing noise produces both $X$ and $Z$ error components, the mapping now requires two sets of gauge spins, $\sigma_j$ for the $X$-stabilizers and $\mu_l$ for the $Z$-stabilizers, with $Y$ errors entering both types.

Following the same procedure, the error distribution $P(I) = 1-p$, $P(X) = P(Y) = P(Z) = p/3$ yields a two-type paramagnet with an $X$-$Z$ cross term~\cite{Bombin2012,chubb2021statmech}:
\begin{equation}
    H_0(\boldsymbol{\tau}^X, \boldsymbol{\tau}^Z) = -J \sum_{q=1}^{n} \left( \tau_q^X + \tau_q^Z + \tau_q^X \tau_q^Z \right),
    \label{eq:depol_paramagnet}
\end{equation}
where $\tau_q^X = (-1)^{e^X_q}$ and $\tau_q^Z = (-1)^{e^Z_q}$. 
The three terms assign energy $-3J$ to the no-error configuration $(\tau^X, \tau^Z) = (1, 1)$ and $+J$ to each of the three single-qubit Pauli errors: $(-1, 1)$ for $X$, $(1, -1)$ for $Z$, and $(-1, -1)$ for $Y$. Each single-qubit error thus introduces a uniform energy penalty $\Delta E = 4J$ relative to the no-error ground configuration. The Boltzmann ratio $P(I)/P(X) = (1-p)/(p/3) = e^{\beta\Delta E}$ then fixes the modified Nishimori condition
\begin{equation}
    \beta J = \tfrac{1}{4}\ln\frac{3(1-p)}{p}.
    \label{eq:depol_nishimori}
\end{equation}

Applying the gauge transformation separately to each channel as in Eq.~\eqref{eq:gauge_mapping},
\begin{equation}
    \tau_q^X = \eta_q^X \prod_{j=1}^{r^X} \sigma_j^{S^X_{j,q}}, \qquad
    \tau_q^Z = \eta_q^Z \prod_{l=1}^{r^Z} \mu_l^{S^Z_{l,q}},
    \label{eq:depol_gauge}
\end{equation}
where $\eta_q^X = (-1)^{\epsilon^X_q}$ and $\eta_q^Z = (-1)^{\epsilon^Z_q}$ are the quenched disorders set by the $X$- and $Z$-components of the pure error, and $S^X$, $S^Z$ are the parity-check matrices of $X$-type and $Z$-type stabilizers, respectively. Substituting Eq.~\eqref{eq:depol_gauge} into Eq.~\eqref{eq:depol_paramagnet} yields the joint Hamiltonian. For the toric code it takes the explicit form
\begin{align}
    H_{\mathrm{toric}}(\boldsymbol{\sigma}, \boldsymbol{\mu}) = -J \sum_{q=1}^{n} \Big[
    &\underbrace{\eta^X_q\, \sigma_{i(q)} \sigma_{j(q)}}_{X\text{ type}} + 
    \underbrace{\eta^Z_q\, \mu_{u(q)} \mu_{v(q)}}_{Z\text{ type}} \notag \\
    &+ \underbrace{\eta^X_q \eta^Z_q\, \sigma_{i(q)} \sigma_{j(q)} \mu_{u(q)} \mu_{v(q)}}_{Y\text{ cross term}} \Big],
    \label{eq:depol_toric}
\end{align}
where $\{i(q), j(q)\}$ label the two vertices ($X$-type stabilizers) adjacent to $q$, and $\{u(q), v(q)\}$ label the two plaquettes ($Z$-type stabilizers) adjacent to $q$. The gauge spins $\{\sigma\}$ and $\{\mu\}$ thus live on the primal and dual lattices, respectively. The $X$ and $Z$ types each take the form of the bit-flip RBIM in Eq.~\eqref{eq:toric_hamiltonian}, while the per-qubit four-body cross terms couple them.

For the $6.6.6$ color code, each bulk qubit lies in the support of three stabilizers per type, so each channel contributes a three-body term as in Eq.~\eqref{eq:color_hamiltonian} (reduced to two-body for boundary qubits), and the $Y$ cross term, as the product of the two channels, couples them through a six-body interaction in the bulk.

\subsubsection{Phenomenological noise model}
\label{sec:phenomenological}

The bit-flip and depolarizing formulations above capture spatial error configurations under perfect syndrome measurements. In practice, syndrome extraction is itself faulty, and the phenomenological noise model therefore requires $T$ consecutive rounds to separate data and measurement errors~\cite{Dennis2002,Fowler2012,Wang2003}. Denoting the set of data qubits by $\mathcal{Q}$ and indexing $Z$-type and $X$-type stabilizer generators by $s \in \mathcal{S}^Z$ and $s \in \mathcal{S}^X$, we record the outcome of stabilizer $s$ at round $t$ as $m_{s,t} \in \{0,1\}$ and convert the raw record into \emph{detection events} $d_{s,t} = m_{s,t} \oplus m_{s,t-1}$, with the initial reference $m_{s,-1} \equiv 0$ given by code initialization in the $+1$ eigenspace of all stabilizers. In the resulting 3D spacetime detector graph, data errors form spatial edges (within a time layer), while measurement errors form temporal edges (connecting adjacent layers)~\cite{Dennis2002,Ohzeki2009,kovalev_spin_2015}.

To accommodate this 3D structure, the variables of Secs.~\ref{sec:bit_flip} and~\ref{sec:depolarizing} are extended, with $t = 0, \dots, T-1$ throughout. The stabilizer gauge spins acquire a temporal index, 
\begin{equation}
\sigma_{s,t} \in \{\pm 1\},\; s \in \mathcal{S}^X; \quad \mu_{s,t} \in \{\pm 1\},\; s \in \mathcal{S}^Z.
\end{equation}
To encode the propagation of data errors across time layers, we further adapt the time-like stabilizer gauge spins~\cite{takada2024colorising} to the codes considered here:
\begin{equation}
\xi_{q,t}^X,\, \xi_{q,t}^Z \in \{\pm 1\}, \quad q \in \mathcal{Q}.
\end{equation}
These couple to the space-like gauge spins so that a data error on qubit $q$ between rounds $t-1$ and $t$ flips the sign of the gauge-invariant data-term product. The initial layer is fixed by $\xi_{q,-1}^X = \xi_{q,-1}^Z \equiv +1$, entering the $t=0$ data terms as constants.

Following the class-by-class framework defined by Eqs.~\eqref{eq:error-configuration} and~\eqref{eq:MWD}, the quenched disorder splits into data and measurement components. To ensure each Hamiltonian corresponds to a single logical class, the label $\boldsymbol{\ell}$ is absorbed into the data quenched disorder:
\begin{equation}
\eta_{q,t}^{(\mathrm{d}),X}(\boldsymbol{\ell}) = (-1)^{\epsilon^{(\mathrm{d}),X}_{q,t} \oplus \delta_{t,0} (\boldsymbol{\ell} L)_{q}^X}, \qquad
\eta_{s,t}^{(\mathrm{m})} = (-1)^{\epsilon^{(\mathrm{m})}_{s,t}},
\label{eq:phenom_quench}
\end{equation}
where $\epsilon^{(\mathrm{d}),X}_{q,t}$ and $\epsilon^{(\mathrm{m})}_{s,t}$ are the data and measurement pure errors consistent with the syndrome (any such pair suffices, as differences correspond to stabilizer elements absorbed by gauge spins), and $(\boldsymbol{\ell} L)_q^X \in \{0, 1\}$ is the $X$-type component of the logical operator at qubit $q$.
Because the logical operator is time-independent, its contribution to the data-error configuration vanishes for $t>0$, restricting it to the $t=0$ boundary. For brevity, the $\boldsymbol{\ell}$ dependence of $\eta^{(\mathrm{d}),X}$ is omitted hereafter.

We begin with phenomenological bit-flip noise, formulating the single-channel Hamiltonian for $X$ errors (detected by $Z$-type stabilizers); the $Z$-channel follows by symmetry. The full Hamiltonian is assembled from three components:
\begin{equation}
H_{\text{phen}}^{X} = H_{\text{data}} + H_{\text{meas}} + H_{\text{final}}.
\label{eq:phenom_total}
\end{equation}

The \emph{data term} penalizes data errors between adjacent time layers, 
\begin{equation}
H_{\text{data}} = -\sum_{t=0}^{T-1}\sum_{q\in\mathcal{Q}} J_{\mathrm{d}}\, \eta_{q,t}^{(\mathrm{d}),X}\, \xi^X_{q,t}\, \xi^X_{q,t-1}\, \prod_{s\in N^X(q)} \sigma_{s,t},
\label{eq:phenom_data}
\end{equation}
where $N^X(q)$ is as defined in Eq.~\eqref{eq:rbim_general}. 

The \emph{measurement term} compares the true stabilizer parity $\prod_{q\in\partial s}\xi^X_{q,t}$ against the quenched disorder $\eta^{(\mathrm{m})}_{s,t}$ from detection events, penalizing mismatches that correspond to measurement errors:
\begin{equation}
H_{\text{meas}} = -\sum_{t=0}^{T-1}\sum_{s\in\mathcal{S}^Z} J_{\mathrm{m}}\, \eta_{s,t}^{(\mathrm{m})}\, \prod_{q\in\partial s}\xi^X_{q,t},
\label{eq:phenom_meas}
\end{equation}
where $\partial s$ is the support of stabilizer $s$. 

The \emph{final-round boundary term} models a perfect final measurement round by pinning every stabilizer's accumulated $\xi$-parity to $+1$ at the final layer,
\begin{equation}
H_{\text{final}} = -J_{\mathrm{s}} \sum_{s\in\mathcal{S}^Z} \prod_{q\in\partial s}\xi^X_{q,T-1} \qquad (J_{\mathrm{s}} \gg J_{\mathrm{d}},\, J_{\mathrm{m}}),
\label{eq:phenom_final}
\end{equation}
where $J_{\mathrm{s}}$ enforces a perfect final syndrome extraction that closes the detector graph at $t = T-1$. 

Let $p$ and $p_{m}$ denote the data error rate per qubit per round and the measurement error rate per stabilizer per round, respectively. The data and measurement couplings satisfy the standard Nishimori relations
\begin{equation}
\beta J_{\mathrm{d}} = \tfrac{1}{2}\ln\frac{1-p}{p}, \quad
\beta J_{\mathrm{m}} = \tfrac{1}{2}\ln\frac{1-p_{m}}{p_{m}}.
\label{eq:phenom_nishimori}
\end{equation}

For this single-channel phenomenological model, each data term contributes $m_{\text{data}} = 2 + |N^X(q)|$ (4-body for the toric code, 5-body for the color code), while the measurement and boundary terms have body counts equal to the stabilizer weight.

For joint depolarizing noise, the full Hamiltonian reads
\begin{equation}
H_{\text{phen}}^{\text{joint}} = \widetilde{H}_{\text{phen}}^X + \widetilde{H}_{\text{phen}}^Z + \widetilde{H}_{\text{phen}}^Y,
\label{eq:phenom_joint}
\end{equation}
where $\widetilde{H}_{\text{phen}}^X$ and $\widetilde{H}_{\text{phen}}^Z$ retain the structural form of Eq.~\eqref{eq:phenom_total} but are evaluated under the joint-channel parameters (measurement errors remain channel-independent), while data errors couple the two channels through a cross term: a $Y$ error corresponds to simultaneous $X$ and $Z$ faults on the same qubit and time layer,
\begin{equation}
\begin{aligned}
\widetilde{H}_{\text{phen}}^Y &= -\sum_{t=0}^{T-1}\sum_{q\in\mathcal{Q}} J_{\mathrm{d}}\, \eta_{q,t}^{(\mathrm{d}),X}\,\eta_{q,t}^{(\mathrm{d}),Z}\, \xi_{q,t}^X\,\xi_{q,t-1}^X\, \xi_{q,t}^Z\,\xi_{q,t-1}^Z \\
&\quad \times \prod_{s\in N^X(q)}\sigma_{s,t} \prod_{s\in N^Z(q)}\mu_{s,t},
\end{aligned}
\label{eq:phenom_Y}
\end{equation}
where $N^Z(q)$ ($N^X(q)$) is the set of $Z$-type ($X$-type) stabilizers acting on qubit $q$, and $\eta^{(\mathrm{d}),X}_{q,t}$ ($\eta^{(\mathrm{d}),Z}_{q,t}$) is the quenched disorder on the $X$ ($Z$) data type, derived from the $Z$-type ($X$-type) stabilizer syndromes. In the joint depolarizing model, the data coupling $J_{\mathrm{d}}$ across all terms in Eq.~\eqref{eq:phenom_joint} follows the joint-channel Nishimori condition, Eq.~\eqref{eq:depol_nishimori}, reflecting the uniform $p/3$ prior for each non-trivial Pauli error; the measurement coupling $\beta J_{\mathrm{m}}$ retains the form of Eq.~\eqref{eq:phenom_nishimori}.

The cross-type $Y$ interaction raises the data-term body count beyond the single-channel count: it involves four time-like stabilizer gauge spins ($\xi^X_{q,t},\, \xi^X_{q,t-1},\, \xi^Z_{q,t},\, \xi^Z_{q,t-1}$) plus $|N^X(q)| + |N^Z(q)|$ space-like stabilizer gauge spins. This yields a local data-term body count of $m_Y = 4 + |N^X(q)| + |N^Z(q)|$, raising the maximum data-term body count to 8-body for the toric code ($|N^X|=|N^Z|=2$) and 10-body for the color code ($|N^X|=|N^Z|=3$).

\section{ITERATIVE LOW-ORDER DECODING ALGORITHM}
\label{sec:ilod}

\subsection{Core mechanism under depolarizing noise}
\label{sec:ilod_depolarizing}

The joint Hamiltonian's $Y$ cross term under depolarizing noise (Sec.~\ref{sec:depolarizing}) has body count $2m$ ($m=2$ for the toric code, $m=3$ for the color code), complicating both the energy landscape and the hardware embedding. To overcome these challenges, we propose an iterative low-order Ising decoding algorithm, which defines the $X$- and $Z$-type sub-Hamiltonians separately and solves them iteratively via a Gauss--Seidel serial schedule~\cite{Saad2003}. The $Y$-type correlations in the joint Hamiltonian are approximated through Bayesian priors that reweight $X(Z)$-type's couplings using the $Z(X)$-type's inferred error configuration. 

The sub-Hamiltonians at the iteration step $\nu$ are defined as follows:
\begin{equation}
\begin{aligned}
H_X^{(\nu)}(\boldsymbol{\sigma}) &= -\sum_{q=1}^{n} J_q^{X,(\nu)}\, \eta_q^X \prod_{j=1}^{r^X} \sigma_j^{S^X_{j,q}}, \\
H_Z^{(\nu)}(\boldsymbol{\mu}) &= -\sum_{q=1}^{n} J_q^{Z,(\nu)}\, \eta_q^Z \prod_{l=1}^{r^Z} \mu_l^{S^Z_{l,q}},
\end{aligned}
\label{eq:ilod_hamiltonians}
\end{equation}
which share the structure of the bit-flip RBIM in Eq.~\eqref{eq:rbim_general} but with qubit-dependent couplings $J_q^{X,(\nu)}, J_q^{Z,(\nu)}$ reweighted at each iteration. Because the $X$- and $Z$-type syndromes are determined by disjoint stabilizer subgroups, the two types carry independent class labels $\boldsymbol{\ell}_X$ and $\boldsymbol{\ell}_Z$. Following the $\boldsymbol{\ell}$-absorption convention of Eq.~\eqref{eq:phenom_quench} (specialized to a single time layer), the quenched disorders $\eta_q^X = (-1)^{\epsilon^X_q \oplus (\boldsymbol{\ell} L)_q^X}$ and $\eta_q^Z = (-1)^{\epsilon^Z_q \oplus (\boldsymbol{\ell} L)_q^Z}$ therefore depend only on $\boldsymbol{\ell}_X$ and $\boldsymbol{\ell}_Z$, respectively. Exploiting this independence, ILOD embeds the class selection directly within each type's update step. In iteration $\nu$, the $X$ type independently evaluates its $2^k$ logical classes (the $X$-type projection of $\boldsymbol{\ell}\in\{0,1\}^{2k}$) under the current couplings $J_q^{X,(\nu)}$. It then selects the minimum-energy class to produce the optimal error estimate $\hat{\boldsymbol{e}}_X^{(\nu)}$, and uses this estimate to update the opposing couplings $J_q^{Z,(\nu)}$ before solving the $Z$ type. For the initial iteration ($\nu=1$), the algorithm uses the unconditioned marginal probability $p_q^{X,(1)} = P(X) + P(Y) = 2p/3$ for every qubit, since no $\hat{\boldsymbol{e}}_Z^{(0)}$ is available for Bayesian conditioning. 

Given the solver's ground state $\boldsymbol{\sigma}_{\mathrm{opt}}^{(\nu)}$ of $H_X^{(\nu)}$, the physical error estimate is recovered by inverting the $X$-type gauge transformation in Eq.~\eqref{eq:depol_gauge}:
\begin{equation}
\hat{e}_{X,q}^{(\nu)} = \tfrac{1}{2}\bigl(1 - \eta_q^X \prod_{j=1}^{r^X} (\sigma_{j,\mathrm{opt}}^{(\nu)})^{S^X_{j,q}}\bigr) \in \{0, 1\},
\label{eq:inverse_map}
\end{equation}
where $\hat{e}_{X,q}^{(\nu)} = 1$ indicates that the solver infers a non-trivial $X$-type symplectic component on qubit $q$ (i.e., an $X$ or $Y$ error), and $0$ indicates its absence. The algorithm then updates the other type's coupling strengths via Bayes' rule. We take the $\hat{\boldsymbol{e}}_X^{(\nu)} \to J_q^{Z,(\nu)}$ update direction as an example (the reverse follows by $X \leftrightarrow Z$ symmetry).

Given $\hat{e}_{X,q}^{(\nu)}$ and treating it as a hard observation of the $X$-type component, Bayes' rule determines the conditional probability that qubit $q$ carries a $Z$ component. If $\hat{e}_{X,q}^{(\nu)} = 1$, the error is restricted to $\{X, Y\}$; the equal priors $P(X) = P(Y) = p/3$ then yield $p_q^{Z,(\nu)} = 1/2$. If $\hat{e}_{X,q}^{(\nu)} = 0$, the error belongs to $\{I, Z\}$ with priors $P(I)=1-p$ and $P(Z)=p/3$; normalization yields $p_q^{Z,(\nu)} = (p/3)/(1-p+p/3) = p/(3-2p)$. Thus,
\begin{equation}
p_q^{Z,(\nu)} = \begin{cases}
\dfrac{1}{2}, & \hat{e}_{X,q}^{(\nu)} = 1, \\[6pt]
\dfrac{p}{3 - 2p}, & \hat{e}_{X,q}^{(\nu)} = 0.
\end{cases}
\label{eq:bayesian_update}
\end{equation}

We assign each $J_q^{Z,(\nu)}$ via the single-type Nishimori form of Eq.~\eqref{eq:paramagnet}, applied locally to $p_q^{Z,(\nu)}$:
\begin{equation}
\beta J_q^{Z,(\nu)} = \tfrac{1}{2} \ln \frac{1 - p_q^{Z,(\nu)}}{p_q^{Z,(\nu)}}.
\label{eq:dynamic_J}
\end{equation}

The alternating optimization proceeds until the estimates $(\hat{\boldsymbol{e}}_X, \hat{\boldsymbol{e}}_Z)$ remain unchanged between consecutive iterations, or until a maximum of $N_{\mathrm{iter}} = 4$ iterations is reached---an empirically determined limit beyond which no further reduction in the logical error rate is observed. The joint logical class $\hat{\boldsymbol{\ell}} = (\hat{\boldsymbol{\ell}}_X, \hat{\boldsymbol{\ell}}_Z)$ from the final iteration determines the decoding output. 

The dynamic and spatially inhomogeneous coupling strengths cause the sub-Hamiltonians to depart from the global Nishimori manifold, as the couplings $J_q^{(\nu)}$ are no longer determined by a single uniform error rate $p$. The alternating procedure is therefore a heuristic optimization guided by local Nishimori relations rather than direct MWD on the joint distribution. The two branches of Eq.~\eqref{eq:bayesian_update} yield distinct local energy landscapes. When $\hat{e}_{X,q}^{(\nu)} = 1$, the resulting $p_q^{Z,(\nu)} = 1/2$ gives $J_q^{Z,(\nu)} = 0$, so the data term indexed by qubit $q$ drops out of $H_Z^{(\nu)}$ entirely. When $\hat{e}_{X,q}^{(\nu)} = 0$, the probability $p_q^{Z,(\nu)} \approx p/3$ lies below the unconditioned marginal $Z$-error rate $2p/3$ that a single-type decoder would use in the absence of Bayesian reweighting, yielding a larger $J_q^{Z,(\nu)}$ that strengthens the penalty against a spurious $Z$ error on that qubit. 

The $X$-$Z$ correlations induced by $Y$ errors are thus approximately captured through the iterative propagation of $\{J_q^{(\nu)}\}$, rather than through the joint high-order terms (explicit 4-body and 6-body cross terms for the toric and color codes, respectively). This Bayesian-reweighting mechanism parallels belief-propagation--reweighted matching decoders~\cite{Criger2018}, which propagate cross-type information through edge-weight updates in matching graphs; ILOD instead propagates correlations through the qubit-dependent couplings of each sub-Hamiltonian.

\subsection{Extension to phenomenological noise}
\label{sec:ilod_phenomenological}

The alternating solve-and-update mechanism extends directly to the phenomenological depolarizing model (Sec.~\ref{sec:phenomenological}), promoting the qubit index $q$ to a spacetime index $(q,t)$, with spacetime gauge spins $\sigma_{s,t}, \mu_{s,t}, \xi^X_{q,t}, \xi^Z_{q,t}$ replacing their static counterparts.

The single-type spacetime sub-Hamiltonian under phenomenological noise comprises data, measurement, and final-round boundary terms. For the $X$ type, it reads
\begin{equation}
\begin{aligned}
H_X^{(\nu)} &= -\sum_{t=0}^{T-1}\sum_{q\in\mathcal{Q}} J_{q,t}^{X,(\nu)}\, \eta^{(\mathrm{d}),X}_{q,t}\, \xi^X_{q,t}\, \xi^X_{q,t-1}\, \prod_{s\in N^X(q)} \sigma_{s,t} \\
&\quad -\sum_{t=0}^{T-1}\sum_{s\in\mathcal{S}^Z} J_{\mathrm{m}}\, \eta^{(\mathrm{m})}_{s,t}\, \prod_{q\in\partial s} \xi^X_{q,t} \\
&\quad -J_{\mathrm{s}} \sum_{s\in\mathcal{S}^Z} \prod_{q\in\partial s} \xi^X_{q,T-1},
\end{aligned}
\label{eq:ilod_phenom_HX}
\end{equation}
where $N^X(q)$, $\partial s$, and $\eta^{(\mathrm{m})}_{s,t}$ are as defined in Sec.~\ref{sec:phenomenological} (the summation $s\in\mathcal{S}^Z$ selects the stabilizer channel that detects $X$-type errors), while the $Z$-type sub-Hamiltonian $H_Z^{(\nu)}$ follows by $X \leftrightarrow Z$ symmetry using variables $\{\mu_{s,t}, \xi^Z_{q,t}\}$.

In the spacetime Bayesian update, only the data couplings $J_{q,t}^{X,(\nu)}$ and $J_{q,t}^{Z,(\nu)}$ are dynamically reweighted; the measurement coupling $J_{\mathrm{m}}$ and the final-round constraint $J_{\mathrm{s}}$ remain static across iterations. This asymmetry reflects the underlying physics, that is, measurement errors are independent classical bit-flips, lacking the cross-type correlations analogous to those produced by Pauli $Y$ errors on data qubits.

The Bayesian update rule of Sec.~\ref{sec:ilod_depolarizing} and the Nishimori mapping, Eq.~\eqref{eq:dynamic_J}, apply unchanged at each spacetime node $(q,t)$, yielding the dynamic couplings $J_{q,t}^{X,(\nu)}$ and $J_{q,t}^{Z,(\nu)}$. 
The inferred data error at each spacetime node is recovered by inverting the gauge-invariant data term
\begin{equation}
\hat{e}_{X,q,t}^{(\nu)} = \tfrac{1}{2}\Bigl(1 - \eta^{(\mathrm{d}),X}_{q,t}\, \xi^{X,(\nu)}_{q,t,\mathrm{opt}}\, \xi^{X,(\nu)}_{q,t-1,\mathrm{opt}} \prod_{s\in N^X(q)} \sigma^{(\nu)}_{s,t,\mathrm{opt}} \Bigr).
\label{eq:phenom_inverse}
\end{equation}
The outer Gauss--Seidel iteration schedule and its convergence criterion remain identical to those in Sec.~\ref{sec:ilod_depolarizing}, applied over the entire spacetime lattice. The global spacetime correlations are handled by the Ising solver within each optimization call, so the $N_{\mathrm{iter}}=4$ iteration limit introduced above remains sufficient across the temporal extents examined in this work.

The joint formulation's $Y$ cross term reaches 8- and 10-body for the toric and color codes, respectively, whereas the corresponding ILOD sub-Hamiltonians retain 4- and 5-body single-type forms. This halves the maximum data-term body count and approximately captures the cross-type correlations through Bayesian-reweighted couplings.

\subsection{Robust refinement under imperfect solvers}
\label{sec:ilod_robust}

The Bayesian derivation assumes that each solver call returns the exact ground state of its sub-Hamiltonian. In practice, the solver may converge to a suboptimal gauge configuration under a finite annealing budget, yielding an imperfect estimate $\hat{\boldsymbol{e}}$ that biases the Bayesian update of $\{J_q^{(\nu)}\}$ and destabilizes the iterative process. To mitigate this, we generalize the hard assignment in Eq.~\eqref{eq:bayesian_update} to a soft-decision formulation via two conditional probabilities reflecting solver reliability:
\begin{equation}
\begin{aligned}  
a &= P(e_{X,q}^{\mathrm{true}} = 1 \mid \hat{e}_{X,q}^{(\nu)} = 1), \\
b &= P(e_{X,q}^{\mathrm{true}} = 0 \mid \hat{e}_{X,q}^{(\nu)} = 0).
\end{aligned}
\label{eq:confidence}
\end{equation}
Marginalizing over $e_{X,q}^{\mathrm{true}}$ via the law of total probability and substituting the ideal conditional probabilities from Eq.~\eqref{eq:bayesian_update} yields
\begin{equation}
p_q^{Z,(\nu)} = \begin{cases}
a \cdot \dfrac{1}{2} + (1 - a) \cdot \dfrac{p}{3 - 2p}, & \hat{e}_{X,q}^{(\nu)} = 1, \\[8pt]
b \cdot \dfrac{p}{3 - 2p} + (1 - b) \cdot \dfrac{1}{2}, & \hat{e}_{X,q}^{(\nu)} = 0.
\end{cases}
\label{eq:robust_update}
\end{equation}
Setting $a = b = 1$ recovers the ideal-solver limit of Eq.~\eqref{eq:bayesian_update}, while $a = b = 1/2$ reduces $p_q^{Z,(\nu)}$ to a constant independent of $\hat{e}_X^{(\nu)}$. 
For practical solvers with $a, b \in (1/2, 1)$, both branches of Eq.~\eqref{eq:robust_update} lie strictly within the interval $(p/(3-2p),\, 1/2)$. Consequently, neither the complete decoupling limit ($J_q^{Z,(\nu)}=0$ for $\hat{e}_{X,q}=1$) nor the strong-penalty limit (for $\hat{e}_{X,q}=0$) is ever reached, preventing erroneous local estimates from propagating as rigid constraints into the opposing type.

The parameters $a$ and $b$ are obtained via offline calibration: for each physical error rate $p$, the conditional probabilities in Eq.~\eqref{eq:confidence} are estimated as empirical frequencies on a calibration set of noise samples with known ground-truth errors, averaged over all qubits and samples. Although the solver reliability in principle varies with the iteration step ($a \to a^{(\nu)}, b \to b^{(\nu)}$), static parameters estimated from the initial iteration ($\nu=1$) suffice to stabilize the updates for all code distances examined here.

\subsection{Efficiency analysis}
\label{sec:ilod_algorithm}

Algorithm~\ref{alg:ilod} summarizes the ILOD procedure, including the sub-Hamiltonian construction (Sec.~\ref{sec:ilod_depolarizing}), spacetime extension (Sec.~\ref{sec:ilod_phenomenological}), and robust refinement (Sec.~\ref{sec:ilod_robust}). By decoupling the joint optimization into alternating types, ILOD reduces the search space, energy-landscape ruggedness, and hardware embedding cost, as analyzed below. \looseness=-1

\emph{Search space reduction.} Each joint solver call explores a gauge space of dimension $2^{r^X + r^Z}$ and enumerates $4^k$ logical classes. ILOD reduces both costs: 1) each ILOD solver call is restricted to a single type of dimension $2^{r^Z}$ or $2^{r^X}$. For the $r^X = r^Z$ codes considered here, this halves the gauge-space exponent, which constitutes the primary search-space saving for the small-$k$ codes examined. 2) In terms of the logical-class enumeration, ILOD evaluates $2^k$ classes per type ($2^{k+1}$ total per iteration) compared to $4^k$ classes in the joint implementation, yielding an additional reduction factor of $\sim 2^{k}$ that would dominate for large-$k$ codes. Beyond the solver calls, the Bayesian coupling updates only introduce linear overhead per iteration, a negligible cost relative to the solver itself. 

\emph{Energy landscape smoothing.} By separating the $X$- and $Z$-type sub-Hamiltonians, ILOD lowers the order of local interactions in the Ising model, which is expected to mitigate the proliferation of metastable states and ease the heuristic ground-state search. Refs.~\cite{derrida1981random,gross1984simplest} establish the analytical bounds of metastable-state proliferation for fully connected spin-glass models; our decoding results on sparse topological graphs are consistent with this landscape-smoothing effect, where ILOD reduces the exponential scaling base of the decoding runtime on the toric code (Figs.~\ref{fig:scaling_cc} and~\ref{fig:scaling_phenom}) and restores convergence at $d=9$ on the color code (Fig.~\ref{fig:color_d9_sweeps}). 

\begin{algorithm}[htbp]
\linespread{1.2}\selectfont
\caption{Iterative low-order decoding}
\label{alg:ilod}
\DontPrintSemicolon
\KwIn{Syndromes/detection events $\boldsymbol{\gamma}^X, \boldsymbol{\gamma}^Z$; physical error rate $p$; number of logical qubits $k$; max iterations $N_{\mathrm{iter}}$; reliability parameters $(a, b)$}
\KwOut{Optimal logical class estimate $\hat{\boldsymbol{\ell}}$}
\BlankLine
Initialize $p_q^{X,(1)} = 2p/3$ for all $q$ \tcp*{unconditioned initial $X$-marginal}
Derive $J_q^{X,(1)}$ via Eq.~\eqref{eq:dynamic_J}\;
\For{$\nu = 1$ \KwTo $N_{\mathrm{iter}}$}{
    \tcp{Phase 1: $X$-type optimization}
    Construct $H_X^{(\nu)}(\boldsymbol{\ell}_X)$ for $\boldsymbol{\ell}_X \in \{0,1\}^k$ using $J_q^{X,(\nu)}$\;
    $\hat{\boldsymbol{\ell}}_X^{(\nu)}, \boldsymbol{\sigma}_{\mathrm{opt}}^{(\nu)}, \hat{\boldsymbol{e}}_X^{(\nu)} \leftarrow$ minimum-energy class and its gauge/error configuration\;
    Update $\{p_q^{Z,(\nu)}\}$ via Eq.~\eqref{eq:robust_update} and derive $\{J_q^{Z,(\nu)}\}$ via Eq.~\eqref{eq:dynamic_J}\;
    
    \tcp{Phase 2: $Z$-type optimization}
    Construct $H_Z^{(\nu)}(\boldsymbol{\ell}_Z)$ for $\boldsymbol{\ell}_Z \in \{0,1\}^k$ using $J_q^{Z,(\nu)}$\;
    $\hat{\boldsymbol{\ell}}_Z^{(\nu)}, \boldsymbol{\mu}_{\mathrm{opt}}^{(\nu)}, \hat{\boldsymbol{e}}_Z^{(\nu)} \leftarrow$ minimum-energy class and its gauge/error configuration\;
    
    \If{$\nu \geq 2$ and $(\hat{\boldsymbol{e}}_X^{(\nu)}, \hat{\boldsymbol{e}}_Z^{(\nu)}) = (\hat{\boldsymbol{e}}_X^{(\nu-1)}, \hat{\boldsymbol{e}}_Z^{(\nu-1)})$}{
        \textbf{break}\;
    }
    
    Update $\{p_q^{X,(\nu+1)}\}$ via Eq.~\eqref{eq:robust_update} and derive $\{J_q^{X,(\nu+1)}\}$ via Eq.~\eqref{eq:dynamic_J}\;
}
\Return $\hat{\boldsymbol{\ell}} = (\hat{\boldsymbol{\ell}}_X^{(\nu)}, \hat{\boldsymbol{\ell}}_Z^{(\nu)})$\;
\end{algorithm}

\emph{Hardware resource reduction.} Mainstream Ising hardware natively supports only 2-body couplings; embedding an $m$-body interaction into quadratic form requires auxiliary spins. Under Rosenberg quadratization~\cite{Rosenberg1975,Boros2002}, depending on the interaction sign and embedding gadget, this decomposition introduces up to $2m - 4$ auxiliary spins per term. ILOD's body count reduction therefore lowers the auxiliary overhead: under phenomenological depolarizing noise, the joint 8-body (toric) and 10-body (color) terms require up to 12 and 16 auxiliary spins per term, while ILOD's 4- and 5-body terms require only 4 and 6 spins per term, respectively.

Combining the above analyses of search-space reduction and per-call solver cost, the total algorithmic cost of ILOD and joint decoding scales as
\begin{equation}
\mathcal{C}_{\mathrm{ILOD}} \approx 2 N_{\mathrm{iter}} \cdot 2^k \cdot C_m, \qquad \mathcal{C}_{\mathrm{Joint}} \approx 4^k \cdot C_{2m},
\label{eq:complexity_comparison}
\end{equation}
where $C_m$ is the cost of approximately finding the ground state of a single $m$-body ILOD sub-Hamiltonian over a gauge space of dimension $2^{r^Z}$ or $2^{r^X}$, and $C_{2m}$ is the corresponding cost for the joint Hamiltonian with $2m$-body interaction terms over a gauge space of dimension $2^{r^X+r^Z}$ (with $m$ identical across the two types of each code).
At fixed $N_{\mathrm{iter}} = 4$ and the small $k$ examined here, the prefactor ratio $2 N_{\mathrm{iter}}/2^k$ favors the joint formulation, so ILOD's empirical advantage stems primarily from the reduced per-call solver cost ($C_m \ll C_{2m}$). For larger $k$, the $2^k$-versus-$4^k$ class enumeration yields an exponential advantage for ILOD.

\begin{figure*}[ht]
    \centering
    \begin{minipage}[t]{0.47\linewidth}
        \centering
        \includegraphics[width=\linewidth]{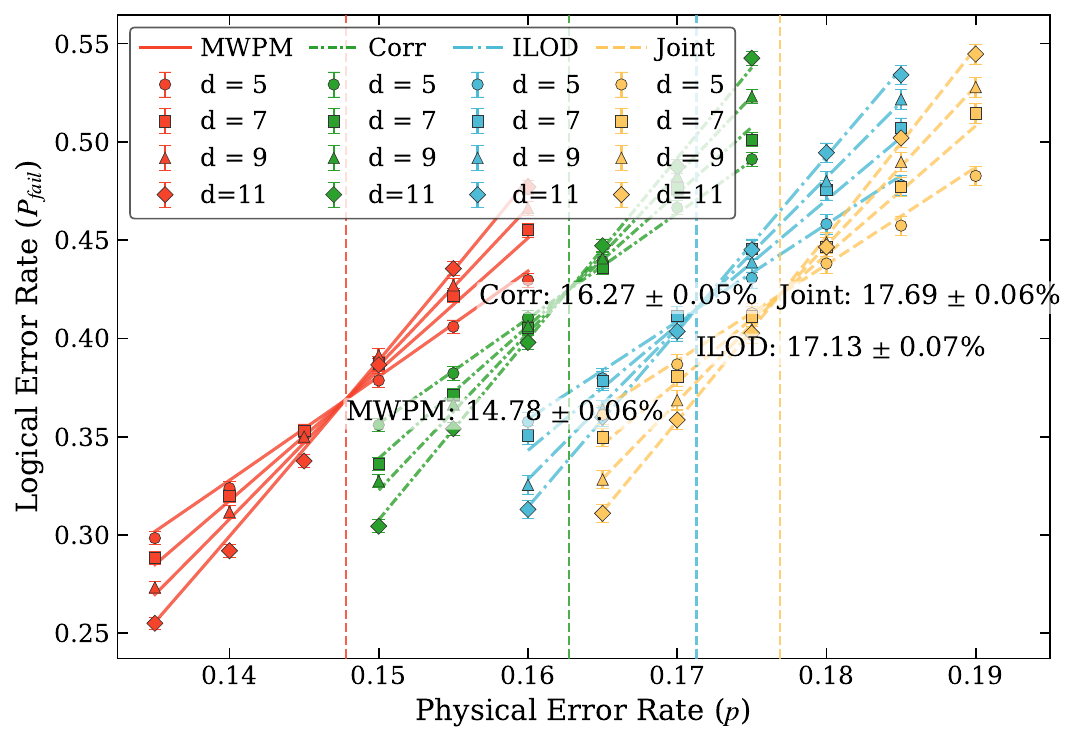}
        \par(a) toric code (code-capacity depolarizing)
    \end{minipage}
    \hfill
    \begin{minipage}[t]{0.47\linewidth}
        \centering
        \includegraphics[width=\linewidth]{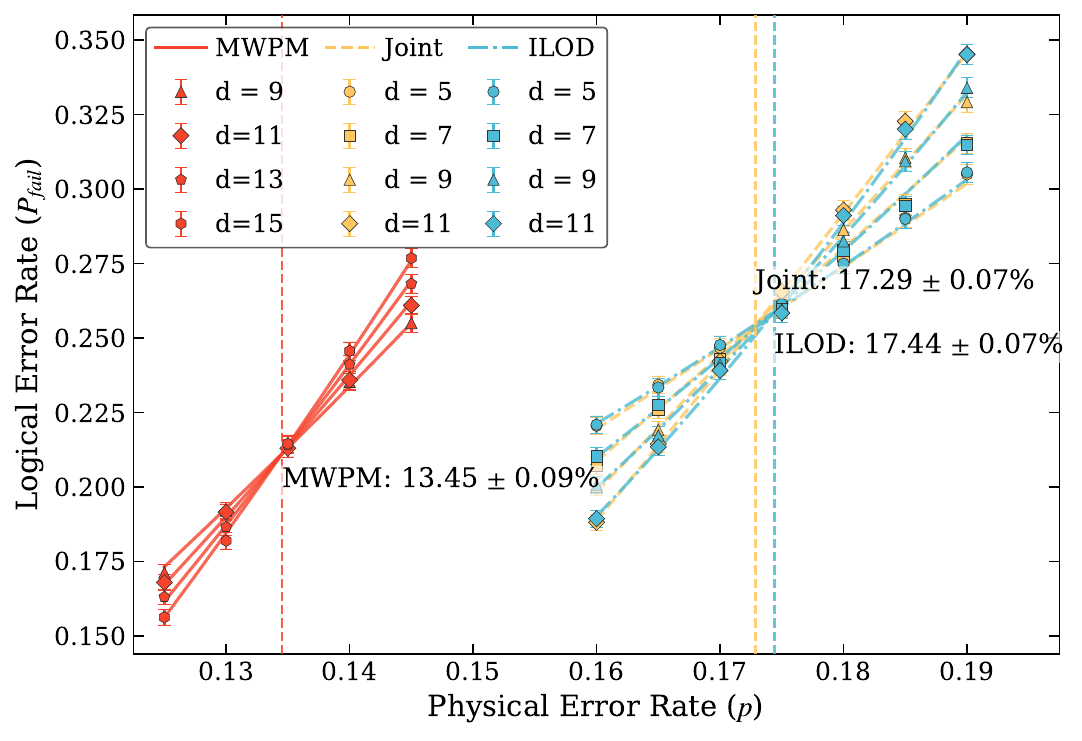}
        \par(b) $6.6.6$ color code (code-capacity depolarizing)
    \end{minipage}
    \caption{Logical error rates under code-capacity depolarizing noise for (a) the toric code and (b) the $6.6.6$ color code. Thresholds are listed in Table~\ref{tab:thresholds}.}
    \label{fig:thresholds_cc}
\end{figure*}

\section{NUMERICAL RESULTS}
\label{sec:results}

This section evaluates ILOD against joint Ising decoding and MWPM in terms of decoding accuracy, runtime scaling, and hardware embedding overhead.
For the matching decoders, we use PyMatching~\cite{higgott2025sparse} for the toric code (standard MWPM and correlated MWPM~\cite{Fowler2013correlated}, denoted Corr-MWPM) and the Chromobius decoder~\cite{gidney2023chromobius} for the $6.6.6$ color code with $X$ and $Z$ syndromes decoded independently. 
For the phenomenological model, we set the measurement error rate equal to the data error rate ($p_{m} = p$) and the number of measurement rounds equal to the code distance ($T = d$)~\cite{Wang2003}. Both the joint and ILOD Hamiltonians are solved by simulated annealing (SA) using the OpenJij library~\cite{openjij} with an exponential cooling schedule ($\beta_{\min}=0.01$ to $\beta_{\max}=10$). To ensure convergence, the sweep count for each $(d, p)$ pair is set to the minimum budget yielding stable logical error rate (LER) estimates within one bootstrap standard error. The ILOD reliability parameters $(a,b)$ (Eq.~\eqref{eq:confidence}) are fixed by offline calibration at each error rate $p$. Each logical error rate is estimated from $2\times10^4$ Monte Carlo trials. Thresholds are extracted using the standard finite-size scaling ansatz $p_L = A + Bx + Cx^2$ with $x=(p - p_{\mathrm{th}}) d^{1/\nu}$~\cite{Wang2003}, via a joint fit across the crossing window.

\begin{table}[htbp]
\renewcommand{\arraystretch}{1.25}
    \centering
    \setlength{\tabcolsep}{4pt}
    \begin{tabular}{@{}lcccc@{}}
        \toprule
        \multirow{2.5}{*}{\textbf{Code}} & \multicolumn{4}{c}{\textbf{Decoding Thresholds ($p_{\mathrm{th}}$)}} \\
        \cmidrule(l){2-5}
         & \textbf{MWPM} & \textbf{Corr-MWPM} & \textbf{Joint} & \textbf{ILOD}\\ 
        \midrule
        \multicolumn{5}{@{}l}{\textbf{\textit{Depolarizing Noise}}} \\
        toric & $14.78(6)\%$ & $16.27(5)\%$ & $17.69(6)\%$ & $17.13(7)\%$ \\
    color & $13.45(9)\%$ & --- & $17.29(7)\%$ & $17.44(7)\%$ \\
        \addlinespace
        \multicolumn{5}{@{}l}{\textbf{\textit{Phenomenological Noise}}} \\
        toric & $3.80(2)\%$ & $4.65(2)\%$ & $4.83(2)\%$ & $4.73(2)\%$ \\
    color & $3.35(5)\%$ & --- & $4.36(5)\%$ & $4.41(4)\%$ \\
        \bottomrule
    \end{tabular}
    \caption{Decoding thresholds for the toric and the $6.6.6$ color codes under code-capacity and phenomenological depolarizing error models. Parenthesized digits give the $1\sigma$ bootstrap uncertainty on the last digit.}
    \label{tab:thresholds}
\end{table}

\subsection{Code-capacity depolarizing noise}
\label{sec:results_cc}

\begin{figure}[htbp]
\centering
    \includegraphics[width=0.96\linewidth]
    {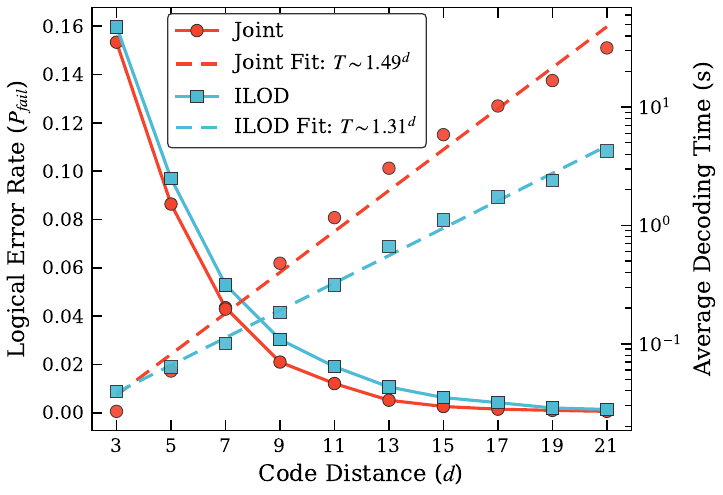}
    \caption{Runtime scaling under code-capacity depolarizing noise for the toric code at $p = 0.10$, fitted as $T_{\mathrm{solve}} \sim b^d$.} 
\label{fig:scaling_cc}
\end{figure}

We first evaluate ILOD under code-capacity depolarizing noise. As shown in Fig.~\ref{fig:thresholds_cc}, the ILOD threshold for the toric code ($17.13\%$) is slightly lower than that of the joint decoding ($17.69\%$). This gap reflects ILOD's approximation penalty, though ILOD still exceeds both MWPM ($14.78\%$) and Corr-MWPM ($16.27\%$), indicating that the Ising formulation treats $X$-$Z$ correlations more accurately than graph matching on these instances. This accuracy penalty is traded for an improved runtime scaling: at $p = 0.10$, the fitted base $b$ of the runtime scaling $T_{\mathrm{solve}} \sim b^d$ drops from $b \approx 1.49$ (Joint) to $b \approx 1.31$ (ILOD), as shown in Fig.~\ref{fig:scaling_cc}.

For the $6.6.6$ color code, the accuracy penalty becomes statistically insignificant: the thresholds of ILOD ($17.44\%$) and the joint decoding ($17.29\%$) agree within $1.5\sigma$ of the combined bootstrap uncertainty. The narrower accuracy gap (compared to $0.56\%$ for the toric code) reflects a closer balance between two competing effects: ILOD's approximation penalty suppresses its own threshold, while the higher body count of the joint Hamiltonian degrades SA convergence at finite sweep budgets and therefore suppresses the measured joint threshold. In comparison, MWPM yields a lower threshold ($13.45\%$), indicating that independent matching of $X$ and $Z$ syndromes does not capture the $X$-$Z$ correlations of depolarizing noise.

\begin{figure*}[htbp]
    \centering
    \begin{minipage}[t]{0.47\linewidth}
        \centering
        \includegraphics[width=\linewidth]{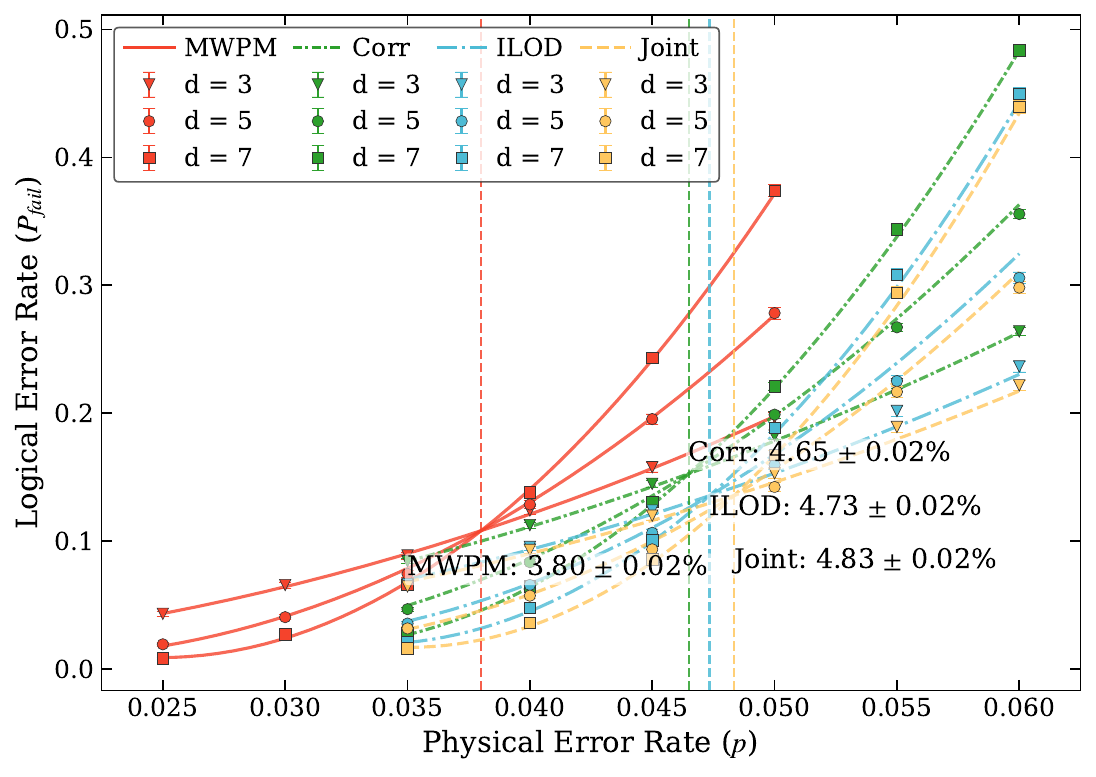}
        \par(a) toric code (phenomenological depolarizing)
    \end{minipage}
    \hfill
    \begin{minipage}[t]{0.47\linewidth}
        \centering
        \includegraphics[width=\linewidth]{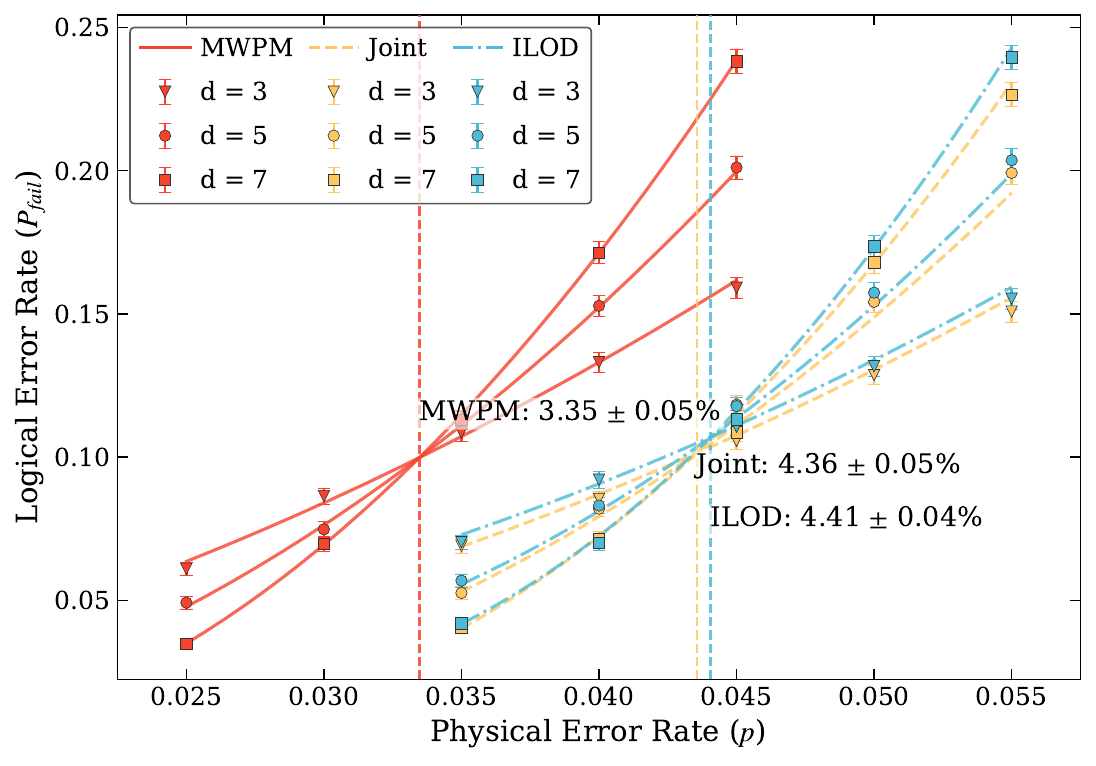}
        \par(b) $6.6.6$ color code (phenomenological depolarizing)
    \end{minipage}
    \caption{Logical error rates under phenomenological depolarizing noise ($p_m=p$, $T=d$) for (a) the toric code and (b) the $6.6.6$ color code. Thresholds are summarized in Table~\ref{tab:thresholds}.}
    \label{fig:thresholds_phenom}
\end{figure*}

\subsection{Phenomenological depolarizing noise}
\label{sec:results_phenom_threshold}

Under phenomenological noise, measurement errors raise the joint Hamiltonian's maximum body count from 4- to 8-body (toric code) and from 6- to 10-body (color code), further complicating the optimization landscape. The two codes behave differently: for the toric code, both Joint and ILOD converge reliably but exhibit different runtime scalings. For the color code, the joint solver converges at $d \leq 7$ but fails to converge at $d=9$ within the sweep budgets examined here, while ILOD remains convergent throughout.

\begin{figure}[htbp]
\centering
\includegraphics[width=0.94\linewidth]
    {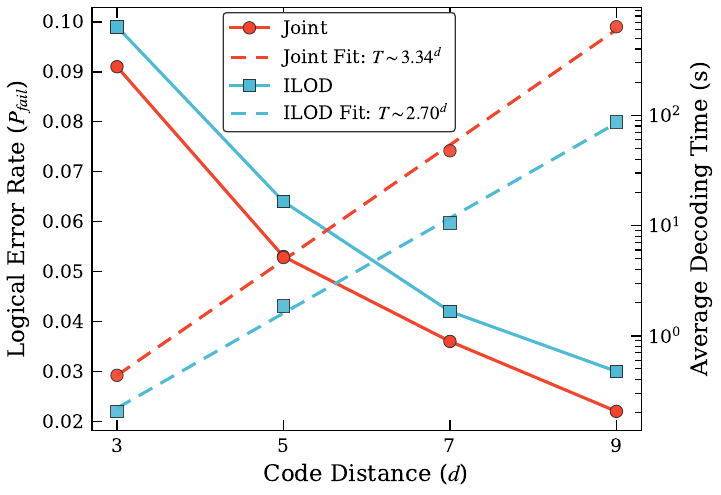}
    \caption{Runtime scaling under phenomenological depolarizing noise for the toric code ($p_m=p$, $T=d$) at $p = 0.03$, fitted as $T_{\mathrm{solve}} \sim b^d$.}
\label{fig:scaling_phenom}
\end{figure}

For the toric code shown in Fig.~\ref{fig:thresholds_phenom}(a), the ILOD threshold ($4.73\%$) is slightly lower than that of the joint decoding ($4.83\%$); both exceed MWPM ($3.80\%$) and Corr-MWPM ($4.65\%$). This threshold gap ($0.10\%$) under phenomenological noise is narrower than the gap ($0.56\%$) observed under code-capacity noise because the 8-body interactions in the joint Hamiltonian further degrade SA convergence, offsetting ILOD's approximation penalty. As quantified in Fig.~\ref{fig:scaling_phenom}, the fitted base $b$ of the runtime scaling $T_{\mathrm{solve}} \sim b^d$ drops from $b \approx 3.34$ (Joint) to $b \approx 2.70$ (ILOD) over $d \in \{3,5,7,9\}$ at $p = 0.03$, yielding an exponentially shrinking runtime ratio with $d$. 

\begin{figure}[htbp]
\centering
\includegraphics[width=0.94\linewidth]{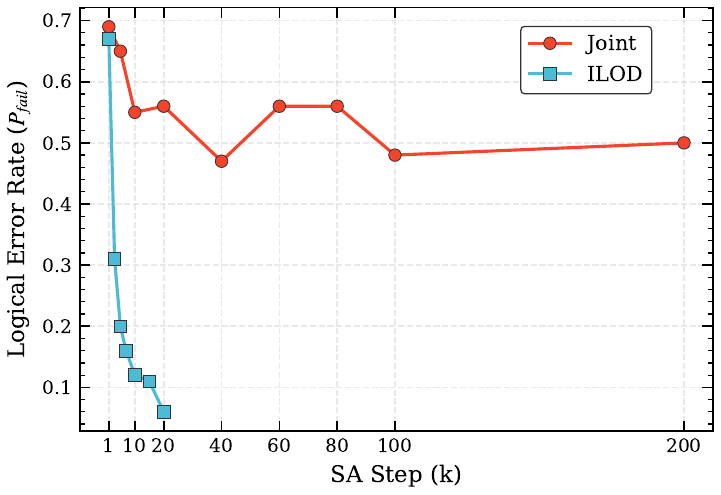}
    \caption{Logical error rate versus SA sweeps per solver call for the $6.6.6$ color code under phenomenological depolarizing noise ($d = 9$, $p = 0.03$, $p_m = p$, $T = d$). ILOD's LER is reported after completing up to $2 N_{\mathrm{iter}} = 8$ sequential sub-Hamiltonian calls per logical class at each sweep budget.}
\label{fig:color_d9_sweeps}
\end{figure}

For the $6.6.6$ color code shown in Fig.~\ref{fig:thresholds_phenom}(b), the thresholds of ILOD ($4.41\%$) and the joint decoding ($4.36\%$) agree within $0.8\sigma$ of the combined bootstrap uncertainty, outperforming MWPM ($3.35\%$). However, for larger distances such as $d = 9$ at $p = 0.03$ (Fig.~\ref{fig:color_d9_sweeps}), the 10-body joint Hamiltonian yields $p_L \approx 0.5$ even at $2 \times 10^5$ annealing sweeps per logical class, with the LER plateauing across the examined sweep range (i.e., SA fails to converge within the available budget). In contrast, ILOD achieves $p_L \approx 0.06$ with only $2 \times 10^4$ sweeps per sub-Hamiltonian solver call, despite the overhead of $N_{\mathrm{iter}}=4$ alternating rounds (two solver calls each).

\subsection{Hardware embedding overhead}
\label{sec:results_hardware}
\begin{figure}[tbp]
\centering
\includegraphics[width=0.95\linewidth]{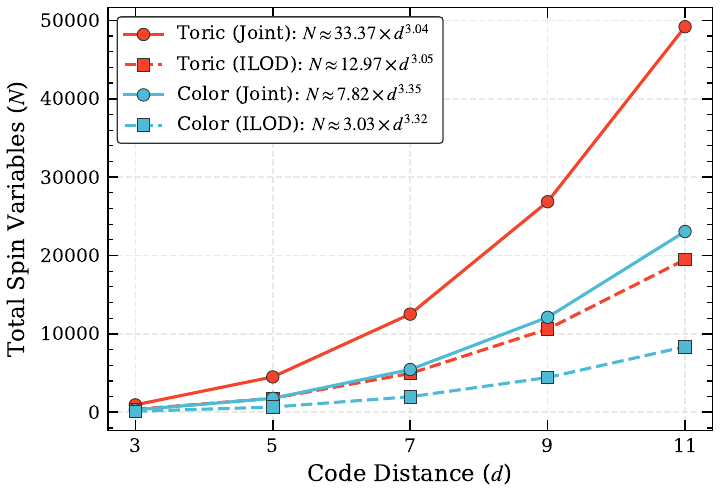}
    \caption{Total spin count (base gauge spins plus auxiliary spins) after Rosenberg quadratization under phenomenological depolarizing noise.}
\label{fig:aux_spins}
\end{figure}

Beyond the accuracy and runtime gains above, ILOD's body-count reduction also lowers the cost of embedding the Hamiltonian into native 2-body Ising hardware. Figure~\ref{fig:aux_spins} reports the total spin count (base gauge plus auxiliary) under phenomenological depolarizing noise after converting the high-order interactions to quadratic form via Rosenberg quadratization~\cite{Rosenberg1975,Boros2002}.
This reduction arises from two effects. First, ILOD discards the high-order cross-terms that dominate the ancilla cost under quadratization. Second, sequential solving of the $X$ and $Z$ sub-Hamiltonians lets the physical spin array be reused across iterations, so the ILOD count reflects the larger of the two type sub-Hamiltonians rather than their sum. 
The resulting spin-count ratio (joint to ILOD) remains nearly constant across the examined distances ($\approx 2.59\times$ for the toric code and $\approx 2.55\times$ for the color code). 

\section{CONCLUSION and DISCUSSION}
\label{sec:conclusion_discussion}

We have proposed ILOD, an iterative Ising decoder that decomposes the joint Hamiltonian under depolarizing noise into single-type ($X$ or $Z$) sub-Hamiltonians, with cross-type information propagated through iterative coupling updates. The numerical results on code capacity and phenomenological noise demonstrate that ILOD 
approximates the threshold of the joint formulation while improving runtime scaling, restoring solver convergence at large code distances where the joint Ising decoder fails to converge, and reducing the total spin count for embedding into native 2-body Ising hardware. Both ILOD and the joint Ising decoder achieve higher thresholds than matching-based methods for the toric and color codes.

Our approach provides an efficient framework for Ising-based decoding.
It can be applied to other CSS codes such as quantum low-density parity-check (qLDPC) codes~\cite{Breuckmann2021,panteleev2021degenerate,Panteleev2022,Bravyi2024} of which non-local connectivity and high stabilizer weights make the embedding overhead of the joint Ising mapping especially severe.
We have considered phenomenological depolarizing noise and the per-qubit prior in ILOD can handle any independent Pauli channel through asymmetric rates $p_x, p_y, p_z$. Extending ILOD to circuit-level noise would require generalizing the per-qubit prior to one that captures inter-qubit correlations, which we leave for future work.
We have used a CPU-based SA solver for numerical evaluation, so the reported runtimes reflect algorithmic scaling rather than the end-to-end latency of a deployed decoder. To achieve lower decoding latency, one can map the Ising Hamiltonians onto dedicated hardware platforms such as coherent Ising machines~\cite{Marandi2014,Inagaki2016}, quantum annealers~\cite{Johnson2011,King2023}, and digital annealers~\cite{Aramon2019}.

\begin{acknowledgments}
We acknowledge the support from QUANTA (\textbf{QU}antum f\textbf{AN}s from I\textbf{T} \textbf{A}rea) group. This work has been supported by the National Key R\&D Program of China (Grant No. 2024YFB4504001), the National Natural Science Foundation of China (Grant No. 62302395, 62301505, and 62421002), the Fundamental and Interdisciplinary Disciplines Breakthrough Plan of the Ministry of Education of China (Grant No. JYB2025XDXM202), the Aid Program for Science and Technology Innovative Research Team in Higher Educational Institutions of Hunan Province. 
\end{acknowledgments}

\bibliography{references}

\end{document}